\documentclass[twocolumn,prl]{revtex4}
\usepackage{graphicx}
\usepackage{amsmath}
\usepackage{amssymb}
\usepackage{bm}
\usepackage{float}
\usepackage{subfigure}
\usepackage{amsfonts}
\usepackage{setspace}
\usepackage[usenames]{color} 
\usepackage{thumbpdf}
\newcommand{\be}{\begin{equation}} \newcommand{\ed}{\end{displaymath}}
\newcommand{\bd}{\begin{displaymath}} \newcommand{\ee}{\end{equation}}
\newcommand{\bea}{\begin{eqnarray}} \newcommand{\eea}{\end{eqnarray}}
\newcommand{\ba}{\begin{array}} \newcommand{\ea}{\end{array}}
\newcommand{\half}{\frac{1}{2}}

\newcommand{\lcdm}{${\rm \Lambda}$CDM}
\begin{document} 
\title{ Detecting Chameleon Dark Energy via Electrostatic Analogy}
\author{Katherine Jones-Smith}
\author{Francesc Ferrer}
\affiliation{Physics Department and McDonnell Center for the Space Sciences, 
Washington University, St Louis, MO 63130, USA}
\begin{abstract} 
The late-time accelerated expansion of the universe could be caused by
a scalar field that is screened on small scales, as in chameleon
or symmetron scenarios.
We present an analogy between 
such scalar fields
and electrostatics,
which allows calculation of the
field profile for general extended bodies.
Interestingly, the field
demonstrates a `lightning rod'
effect, where it becomes enhanced 
near the ends of a pointed or elongated object.
Drawing from this correspondence, we show that non-spherical
test bodies immersed in a background field will experience
a net torque caused by the scalar field.
This effect, with no counterpart in the gravitational case, 
can be potentially tested in future experiments.
\end{abstract}
\maketitle 
Among the host of cosmological observations that the concordance model, \lcdm,
accounts for~\cite{Komatsu:2010fb} the late-time accelerated expansion 
of the universe poses one of the most compelling problems in physics. 
As inferred from different distance measurements, the energy density of
the universe is presently dominated by a  dark energy component with
strongly negative pressure;
in \lcdm\ this role is played by a cosmological constant $\Lambda$. Its extremely small value, at odds with theoretical expectations, has prompted
the exploration of alternative models in which a scalar field $\phi$ 
causes the cosmological acceleration (see 
e.g.~\cite{Copeland:2006wr,Caldwell:2009ix,Silvestri:2009hh}).

To obtain dark energy behavior
with an evolution of the energy density, however, the scalar
must be extremely
light,
and generically mediates
horizon ranged interactions. 
Interestingly, constraints
from fifth force experiments
can be evaded in modified gravity theories, where {\it screening mechanisms}
(reviewed in~\cite{Khoury:2010xi}) 
are at play in dense environments. 
The scalar could then have significant couplings with ordinary matter
at the fundamental level, but the range of the scalar mediated force becomes 
small enough to conform to terrestrial measurements ~\cite{Mota:2006fz}.
The force gets further reduced for relatively large bodies if, 
as
in {\em e.g.} chameleon or symmetron scenarios, it is only sourced over
a thin shell under the surface of the body, thus avoiding conflict with precision
measurements of gravity in the sparser environment of the solar system.

In this Letter, we construct a mathematical analogy between the chameleon scalar field and the electrostatic potential. This analogy is in the same spirit as the many systems identified in the chapter devoted to electrostatic analogs in Feynman's classic text \cite{feynman},  for example, the uniform illumination of a plane,  or the flow of an irrotational fluid past a sphere. The underlying principle at work here is that the same equations have the same solutions---electrostatics 
is not germane to the analogous system, but rather the phrase `electrostatic analogy' serves as proxy for saying that the analogous system obeys the same differential equations as electrostatics.
Of course, not all phenomena from an electrostatic system will have counterparts in the analogous system and vice versa; 
nonetheless, as
Feynman shows, one can fruitfully export intuition and solutions from electrostatics to its
analogs.
We show in this  Letter that under conditions relevant to terrestrial experiments the chameleon obeys
the same equations as the electrostatic potential.
Our central finding is that the chameleon field outside of elongated bodies such as ellipsoids is enhanced relative to the spherical bodies typically considered. This shape enhancement can be exploited by experimenters to probe new regions of chameleon parameter space, even in the experimentally unfavorable thin shell regime.

Fields featuring screening mechanisms 
often appear in scalar-tensor modifications of general relativity. 
We choose
a chameleon field $\phi$ 
to exemplify our calculations ~\cite{Khoury:2003aq,Khoury:2003rn}, but our results extend
to other models, like the 
symmetron~\cite{Pietroni:2005pv,Olive:2007aj,Hinterbichler:2010es}, 
which also exhibit the thin shell.  
In these models, matter follows the geodesics of the metric $\tilde{g}_{\mu \nu}=
A^2(\phi) g_{\mu \nu}$, where
$\phi$ is a scalar field. Due to this conformal coupling the effective potential that
appears in the Klein-Gordon equation includes a piece that depends on
$\rho_m$
the density of matter,$V_{eff}(\phi)=V(\phi)+ A (\phi) \rho_m$.   In the chameleon family of models $V(\phi)$ is monotonically decreasing while $A(\phi)$ is
monotonically increasing~\cite{Khoury:2003aq,Khoury:2003rn}. With this 
choice the effective potential has a minimum;
both the value of the field at the minimum and the mass of the field,
$m_\phi^2=\partial^2 V_{eff}/\partial\phi^2$, 
depend on $\rho_m$.
Thin shells do not exist for all couplings and 
potentials~\cite{Tamaki:2008mf}, so
from now on we make the usual choice
$A(\phi)={\rm e}^{\beta \phi/M_{pl}}$.
Replacing the dimensionless constant
$\beta$ by $\alpha(\phi_\infty)\equiv M_{pl}\;
{\rm d}\log A/{\rm d}\phi$, where $M_{pl} = 1/\sqrt{8 \pi G}$, 
our results also apply to generic conformal couplings.

For static configurations of the field in a weak gravitational background,
the chameleon obeys the Klein-Gordon equation
\be
 \nabla^2\phi=\partial V_{eff}/\partial \phi.
\label{eqnmotion}
\ee
In general this is an unwieldy non-linear differential equation, but in some regimes, approximations allow for its easy solution. 
For example, the chameleon field outside a sphere (radius $R_c$, density $\rho_c$, Newtonian gravitational potential $\Phi_c$ 
) immersed in a uniform background medium (density $\rho_\infty < \rho_c$)  is well-approximated by a Yukawa profile~\cite{Khoury:2003aq,Khoury:2003rn}.  There are two versions of the Yukawa profile, depending on 
the density contrast between the body and its environment~\cite{Khoury:2003aq,Khoury:2003rn}. If the density contrast is slight, the body does not disturb the surroundings too much, and the field is only slightly perturbed from the value which minimizes $V_{eff}$ corresponding the density $\rho_\infty$ ( we denote this value $\phi_\infty$).  This is known as the thick shell regime; see ~\cite{Khoury:2003rn} for details. If a chameleon field with the canonical gravitational-strength coupling $\beta \sim 1$ was ambient, bodies with thick shells would experience a chameleon force of roughly the same magnitude as their gravitational force, making detection of the chameleon field easy~\cite{Khoury:2003aq}. However virtually all terrestrial objects are in what is known as the thin shell regime, so we henceforth ignore the thick shell regime.  In the thin shell regime, the density contrast is great enough so as to make the field inside the body virtually impervious to the field outside the body. That is, throughout the bulk of the body (what we will call the core region) the field simply sits at the minimum of $V_{eff}$ appropriate to the density $\rho_c$; we call this value $\phi_c$. It only starts to `see' the exterior density as $r$ approaches $R_c$, and only over the course of a thin shell of material (thickness $\Delta R_c$) just underneath the surface of the body does the field begin to vary. Once outside the body, the field is well-described by a Yukawa profile
\be
\phi(r) \approx \left(\frac{-\beta}{4\pi M_{Pl}}\right)
\left(\frac{3\Delta R_c}{R_c}\right)\frac{M_ce^{-m_{\infty}(r-R_c)}}{r} 
+\phi_{\infty},
\label{thinshell}
\ee  
where
\be
\frac{\Delta R_c}{R_c}\equiv \frac{\phi_\infty -\phi_c}{6\beta M_{Pl}\Phi_c}\ll1
\label{shelldef}
\ee

Note however that under most circumstances of interest (such as a vacuum chamber 
experiment to probe violations of the equivalence principle),
the argument of the exponent is so small that the chameleon profile outside the body is $1/r$ for all intents and purposes: 
\be
\phi=\phi_\infty + (\phi_c-\phi_\infty)\frac{R_c}{r}.
\label{sphereprofile}
\ee
which is the same as ~Eq.~(\ref{thinshell}) as the reader can check. 
To summarize, in the thin shell regime, $\phi$ is constant throughout its core. Only a thin shell of material just underneath the surface contributes to the exterior field, where the profile is technically Yukawa. Given the densities and distances relevant to a terrestrial experiment, we can ignore the exponent, yielding simply a $1/r$ behavior. 

But we can also obtain this $1/r$ behavior in a simpler way: a massive Yukawa profile results from making a second order Taylor expansion of the field around its minimum, reducing Eq.~\ref{eqnmotion} to  $\nabla^2 \phi \approx m^2\phi$. If we find later we can ignore the mass, we may as well just solve Laplace's equation, $\nabla^2 \phi =0$
to determine the chameleon field exterior to the body. The solution to Laplace's equation in spherical coordinates is simply $\phi= A + B/r$ where $A,B$ are determined by boundary conditions. Very far from the sphere we should  have $\phi \rightarrow \phi_\infty$,  and at $r=R_c$ we assume $\phi=\phi_c$, hence $A=\phi_\infty$ and $B=(\phi_c-\phi_\infty)R_c$.  Thus we recover Eq.~(\ref{sphereprofile}). 

Note the collective 
behavior of the chameleon field for a thin shelled sphere is precisely the same as the behavior of the electrostatic potential $\psi$ for a conducting sphere: inside the sphere both $\psi$ and $\phi$ are constant, and outside the sphere both fields obey Laplace's equation. In a conductor there is a thin layer of charge residing on the surface that sources the electric field outside the sphere; similarly inside the thin shell region the chameleon field Eq.~(\ref{eqnmotion}) may be approximated as Poisson's equation, with $\nabla^2 \phi = \beta \rho_c/M_{Pl}$. 

Thus there is an analogy between the chameleon field for thin shelled objects and the electrostatic potential of conducting objects, 
by
the principle that the same differential equations have the same solutions.  

Let us examine more closely the charged surface layer.
We know how to interpret this in an electrostatic context-- we say there is a surface layer of electric charge $\sigma$, which is related to the external field gradient via $\partial \psi/\partial n=\sigma$, where $n$ is the direction normal to the surface (and we have taken $\epsilon_0=1$).  
Similarly we can
write $\partial \phi/\partial n= \varrho \delta$ where $\varrho = \beta \rho_c/M_{Pl} $ is the volume density of `chameleon charge' and $\delta$ is the thickness of the layer over which this chameleon charge is distributed. For the sphere, setting
$\partial \phi/\partial n = (\beta \rho_c/M_{Pl}) \delta$, and solving for $\delta$ we find 
\be
\delta=\frac{(\phi_\infty-\phi_c)R_c}{6\beta M_{Pl}\Phi_c}, 
\ee
which is identical to the thickness of the shell $\Delta R_c$ found in a completely independent way by Khoury and Weltman, Eq.~(\ref{shelldef}). 
This suggests we interpret $\beta \rho_c/M_{Pl}$ as the volume density of `chameleon charge'. Doing so independently reproduces the thickness of the shell derived in \cite{Khoury:2003rn}, and is consistent with Eq.~(\ref{eqnmotion}) reducing to Poisson's equation inside the shell region with the RHS given by $\beta \rho_c/M_{Pl}$.  Note that this chameleon charge represents the material within the body that interacts with the chameleon field outside, and that it is confined to the shell/surface layer.  Note further that the the thick shell regime does not extrapolate from the thin shell regime increasing the thickness of the shell  (in fact it doesnt extrapolate at all). 
 
Given that the thin shell effect arises via the density contrast and boundary conditions, it stands to reason that although the effect was first derived for a sphere, {\em ceteris paribus} a less symmetric shape would still possess a thin shell. We can use the analogy to determine the chameleon profile for less symmetric shapes; in  this Letter we work with ellipsoids to illustrate the interesting shape-dependent effects we have identified.
Ellipsoids also have the merit that they 
can also be compared with spherical results in the limit that the eccentricity $\varepsilon \rightarrow 0$.

 Ellipsoids are described by the prolate spheroidal coordinates 
 $(\xi, \eta,\varphi)$; the surface of an
 ellipsoid has radial coordinate $\xi=\xi_0$; 
 furthermore $\xi=1/\varepsilon$. 
 $\eta$ measures the latitude, 
 with the poles at $\eta = \pm 1$ and the equator at  $\eta = 0$.  
 It is convenient 
 to introduce
 an equivalent radius $R_e$ such that the volume of the 
 ellipsoid is $\frac{4}{3} \pi R_e^3$.
 We first consider the chameleon field {\em produced} by an ellipsoid of arbitrary material, assuming only that it possesses a thin shell. Its interior is at a constant value determined by the density $\rho_c$, and in the exterior it is the solution to Laplace's equation. The relevant solution to Laplace's equation in these coordinates is obtained from \cite{mandf};
 \be
 \phi=
      \phi_\infty +(\phi_c -\phi_\infty)\frac{Q_0^0 (\xi)}{ Q_0^0 (\xi_0)} 
 \label{psiprof2}
 \ee
 where $Q_0^0(\xi)= \ln[(\xi+1)/(\xi-1)]/2$ and we have used appropriate boundary conditions. 
 Assuming $r \gg a$ where $a$ is the interfocal distance of the ellipsoid and $r$ is the radial spherical coordinate, the chameleon profile can be written
 \be
 \phi = \phi_\infty - f(\xi_0)(\phi_\infty - \phi_c)\frac{R_e}{r}
 \label{chamellipse}
 \ee
 where \be
 f(\xi_0)=\frac{2}{[\xi_0(\xi_0^2-1)]^{1/3}}\frac{1}{\ln[(\xi_0 +1)/(\xi_0 -1)]} ,
 \label{f}
 \ee
 and we have set  $a=2R_e/[\xi_0(\xi_0^2-1)]^{1/3}$. 

In comparing to the case of the sphere (Eq.~\ref{sphereprofile}) we see the ellipsoid has a shape enhancement $f(\xi_0)>1$ which diverges as the ellipsoid flattens to a line. To see the effect this shape enhancement might have on experiment, consider the following Cavendish-type set up.  A test mass is located some distance from the pole of a spherical source body. There is of course a  gravitational attraction between the source and the test mass; if the chameleon field exists, 
there will also be a `fifth' force exerted on the test mass by the chameleon field of the source. 
 The chameleon force on a test mass $m$ is given by \cite{Khoury:2003rn}
 \be
{\mathbf F}_5 = - m \nabla ( \beta \phi/ M_{Pl}). 
\label{testmass}
\ee
Using the chameleon profile outside the sphere, Eq.~\ref{sphereprofile}, there will be unmitigated suppression of the force via the thin shell factor. 
However, if we were to use an ellipsoidal source instead of a spherical one, we would use Eq.~\ref{chamellipse} instead of Eq.~\ref{sphereprofile} in calculating the force. The ellipsoidal profile still possesses the thin shell suppression factor, but the suppression effect can leveraged by the enhancement $f(\xi)$. 

We refer to this mitigation of the suppression factor as a 'lightning rod' effect because in electromagnetism the electric field at the polar region of an elongated object is enhanced relative to the polar region of a sphere,  though we stress once again that electromagnetic phenomena are not germane to this set up. The enhancement arises as a feature of the elongated ellipsoid, which has a preferred axis (its major axis) which the sphere lacks. The reader can confirm that as $\varepsilon\rightarrow 0$, the spherical results are obtained. An experiment would likely probe the near field close to the sharp tip of a dense body rather than the asymptotic field discussed above. We will return to the analysis of realistic force experiments in future work. 
 %
 %
 %
 
 We now turn to another shape enhancement
 effect demonstrated by the chameleon field. We begin by calculating the chameleon force on an extended body that cannot be treated as a test mass. It follows from Eq.~\ref{testmass} that the force on an extended body with density $\rho_c$ and volume element $dV$ be given by 
 \be
 {\mathbf F} = \int_{{\rm vol}} d V \; \frac{ \beta \rho_c }{M_{Pl}} \nabla \phi.
 \label{eq:exactexpression}
 \ee
In the thin shell regime $\phi=\phi_c$ in the core so it is
only necessary to integrate over the shell.  As noted the field obeys Poisson's equation in this region. We denote the thickness of the shell as $\Delta R$, and take $z$ to be the coordinate along the local normal to the surface $\hat{{\mathbf n}}$.  Assuming the gradient of the field vanishes at $z=0$ where the shell meets the core, we obtain 
 \be
 \phi = \phi_c +  \frac{1}{2} \frac{\beta \rho_c}{M_{Pl}} z^2.
 \label{eq:shellprofile}
 \ee
 Substituting eq (\ref{eq:shellprofile}) into eq (\ref{eq:exactexpression}) yields
 \begin{equation}
 {\mathbf F}_{{\rm net}} = \int d a \; \int_0^{\Delta R} d z \; \left( \frac{\beta \rho_c}{M_P} \right)^2 z \; \hat{{\mathbf n}}
 = \frac{1}{2} \int d a \; \left( \frac{ \beta \rho_c \Delta R }{M_{Pl}} \right)^2 \hat{{\mathbf n}}.
 \label{eq:integralevaluated}
 \end{equation}
 Using $\beta \rho_c \Delta R/M_{Pl} = \partial \phi/\partial n$ we can write this as
 \be
 {\bf F}_{{\rm net}} = \frac{1}{2} \int d a \; \left( \frac{ \partial \phi }{ \partial n } \right)^2 \hat{{\mathbf n}}. 
 \label{thinforce}
 \ee
 Note that this is the same expression for the force on a conductor in an electrostatics context, if we take $\epsilon_0 = 1$ and replace $\phi$ by the electrostatic potential $\psi$.
Using Eq.(\ref{thinforce}) one can calculate the chameleon force on a spherically symmetric extended body and show that it only differs from that of the test mass by the thin shell factor. At first this may seem surprising, because it says a test mass experiences greater force than the extended object of the same mass. But the entire test mass experiences the chameleon field gradient whereas in the extended body, only the material in the shell sees a field gradient, so the suppression makes sense. In graduating from a test mass to an extended sphere, the body `acquires' a thin shell. 

So the chameleon field exerts a force on an extended body given by Eq.~\ref{thinforce}.  It follows that if  this force acts at a distance ${\mathbf r}$ it will result in a torque, given by ${\boldsymbol \tau} = {\mathbf r} \times {\mathbf F}$:
 \be
 {\boldsymbol \tau} = \half \int d a \; (\partial \phi / \partial n)^2 
 {\mathbf r} \times \hat{{\mathbf n}}.
\label{chamtorqe}
\ee
Note that an extended body can experience a torque, where a true test mass cannot--treated as a point particle it has no radial extent and thus lacks a lever arm at which the force can act to produce a torque. Combining this result with our earlier finding that the chameleon field {\em produced} by an ellipsoid is enhanced in the polar regions, we suspect an ellipsoid immersed in a chameleon field will experience non-zero torque, due wholly to the chameleon field.  
To determine this we need to pull together several results.  
Khoury and Weltman \cite{Khoury:2003rn} argue that the ambient chameleon field inside a vacuum chamber has a uniform gradient, 
whose magnitude $\chi=|\nabla \phi|$ they estimate. Introducing a test mass would not disrupt this field configuration but since we are considering an extended source we seek the solution to Laplace's equation in which the ambient chameleon field adjusts to the presence of this extended body, which is also a source of chameleon field. 
Morse and Feshbach~\cite{mandf} provide the solution to Laplace's equation for a conducting ellipsoid whose interior is held at a constant potential and which is immersed in a uniform electric field that makes an angle $\gamma$ with the ellipsoid's major axis; by analogy the chameleon field for a thin shelled ellipsoid immersed in a chameleon field with uniform gradient is given by the same expression. The resulting expression for the chameleon field is $\phi_1+\phi_2$ where $\phi_1$ is given by Eq.~\ref{psiprof2}) and 
 \bea
 \phi_2 & = & \chi \frac{R_e}{[ \xi_0 ( \xi_0^2 - 1 ) ]^{1/3}} \{ \cos \gamma \; \eta \left[ \frac{ \xi_0 }{ Q_1^0 (\xi_0) } Q_1^0 (\xi ) - \xi \right] \nonumber \\
 & + &  \sin \gamma
 \cos \varphi  \sqrt{1 - \eta^2}
 \left[ 
 \frac{\sqrt{ \xi_0^2 - 1 } }{Q_1^1 (\xi_0) } Q_1^1 (\xi) - \sqrt{ \xi^2 - 1 } \right] \}
 \nonumber \\
 \label{eq:psi2}
 \eea
 Here $Q_0^1 (\xi )  = \xi Q_0^0 (\xi) - 1$ and 
 $Q_1^1 (\xi) = \sqrt{ \xi^2 - 1} [ \xi/( \xi^2 - 1 ) - Q_0^0 (\xi) ]$. 

For this profile, the chameleon torque on the ellipsoid has one non-vanishing component 
 \be
 \tau_y = \pi R_e^3 \chi^2 \sin \gamma \cos \gamma\; g(\xi_0)
 \label{eq:torque}
 \end{equation}
 where the shape dependent factor 
 \be
 g( \xi_0 ) = \frac{ 2/3 + 2 ( 1 - \xi_0 ) Q_1^0 (\xi_0) }{ Q_1^0 (\xi_0) Q_1^1 (\xi_0)
 \xi_0^{5/3} ( \xi_0^2 - 1 )^{3/2} }.
 \label{eq:shapefactor}
 \ee
 Note that the torque vanishes in 
 the spherical limit 
 $ \xi_0 \rightarrow \infty$.
It is remarkable that the chameleon field produces a torque because a gravitational field with uniform gradient
 would not produce a torque on an ellipsoid\footnote{
 To the extent that both fields are uniform, the torque is due only to the chameleon. In a real experiment of course neither field would be perfectly uniform so there would be corrections to this.}. 
 Before estimating the magnitude of the torque we point out several important features. $\tau$ is independent of $\beta$ and $\rho$; this is analogous to the corresponding electrostatic torque on a conductor being independent of the total charge.  The insensitivity of terrestrial experiments to $\beta$ is well-known~\cite{Mota:2006fz} and unfortunate from the stand point of detection. However, any value of $\beta \gtrsim1$ leads to the formation of a thin shell for terrestrial objects, hence if  the torque experiment were to yield a null result (and all other experimental factors are accounted for), this entire range of $\beta$ would be ruled out.   
 To estimate the magnitude of the torque we follow \cite{Khoury:2003rn} and use
 $V(\phi)=M^5/\phi$,
 $R_{vac} \sim 1$ m and $M \sim 10^{-3}$ eV. A reasonable choice of ellipsoid is  $R_e = 0.1$ m, so that $R_e \ll R_{vac}$; this ensures the gradient of the chameleon seen by the ellipsoid is uniform. We choose $\xi_0 = 1.3$ 
 so that the ellipsoid is distinctly non-spherical but not extremely elongated. With these values we find that the torque $\sim 7 \times 10^{-15}$ Nm. This is promising as the 
 E\"ot-Wash experiments are sensitive to torques of this magnitude~\cite{Kapner:2006si},  so an experiment with similar sensitivity could rule out $\beta >1$. Parenthetically we note that using an ellipsoid constructed from electrically insulating material in an actual experiment would be desirable for suppressing spurious electrostatic torques.

 The authors are grateful to Harsh Mathur for useful conversations.  This
 work was supported in part by the U.S. DOE under Contract
 No. DE-FG02-91ER40628 and the NSF under Grant No. PHY-0855580. 

\bibliography{chameleon_v6}

\end{document}